\documentclass [a4paper,12pt]{article}
\usepackage{amsmath,amssymb,graphicx,cite,delarray}
\usepackage[a4paper]{geometry}
\usepackage[bf]{caption2}

\begin{document}
\pagestyle{plain}
\begin{center}\begin{Large} Analogy between the wave of the falling dominoes 
and the growth of martensitic crystal (a simple enunciating for experimenters)
\end{Large}\end{center}

\begin{center}
M.P.\,Kashchenko
\end{center}

Physics Chair, Ural State Forest Engineering University, 
Sybirskiy trakt, 37, 620100, Ekaterinburg, Russia

\begin{abstract}A number of laws being characteristic for switching waves are illustrated by the examples of waves of a falling dominoes. The specificity of a switching wave at the martensite crystal growth caused by dynamic structure of interphase area is noted. For the first time the influence of finite deformation on the condition of elastic waves generation by non-equilibrium electrons is discussed. The rigid regime of initial excitation of waves is connected with the influence of finite deformations on the threshold value of an inverse population difference of the electronic states.
\end{abstract}


\section{INTRODUCTION}

The wave phenomena are manifold. For example oscillations with small amplitudes may spread in any mediums. Velocity of waves depends on interaction of the structural elements of medium. Further the regular arrangement of the structural elements in space is supposed. In particular, atoms in crystals form a periodic spatial lattice and a spread of waves with small amplitudes is accompanied by small displacements of atoms from standings of equilibrium. If the structural elements have some positions of equilibrium with the various reserved energy the motion is possible at which building blocks consistently transfer from the position with the greater energy to the position with smaller energy. The wave of falling dominoes is an example of such waves of switching. The comprehension of the features of this wave is useful for the interpretation of the martensite crystal growth. These features are listed in a kickoff of the report.

Then the features of modified switching waves controlling the martensite crystal growth are considered. In a final part the new results are given. These results are essential to improvement of the conditions of wave generation at considerable strains breaking symmetry of a lattice.

\section{SWITCHING WAVE IN DOMINO SYSTEM}
\textbf{A necessary condition for existence of a  switching wave.}

\begin{figure}[htb]
\centering
\includegraphics[clip=true, width=0.8\textwidth]{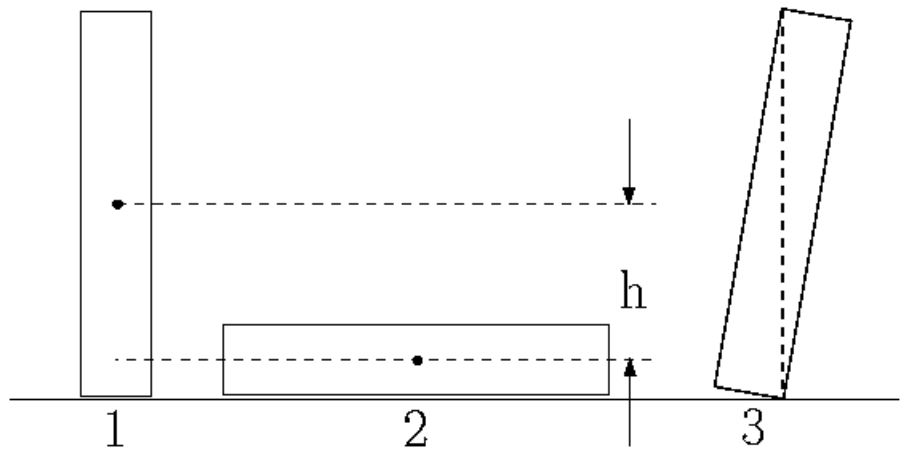}
\caption{The equilibrium positions of a domino.}
\label{fig1}
\end{figure}

\begin{figure}[htb]
\centering
\includegraphics[clip=true, width=0.8\textwidth]{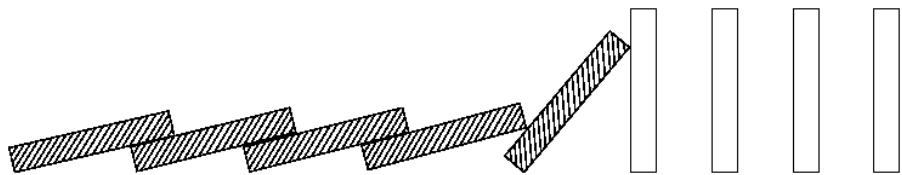}
\caption{The switching wave in a domino.}
\label{fig2}
\end{figure}

The separate domino is an element of medium with three equilibrium positions shown in a fig.~ \ref{fig1}. The first and second positions are metastable against diversions in some intervals of values (smaller interval for the first position) and third position is labilely to as much as small diversions. 
The third position has the greatest potential energy $U_{3}$ in a gravitational field.
The quantity $\Delta U_{12} = U_{1} - U_{2}$ increases as the difference of vertical coordinates of a centre of gravity of a domino. The difference of energies $\Delta U_{31} = U_{3} - U_{1}$ is an energy barrier. The transition $1 \longrightarrow 2$ is impossible without overcoming of this barrier. As it follows from the law of conservation of energy, after overcoming a barrier the difference of potential energies $\Delta U_{32} = U_{3} - U_{2}$ will transferred into a kinetic energy $E$. If the energy $E$ exceeds a barrier $\Delta U_{31}$ there is an opportunity for self-sustaining (or "autocatalytic") of switching wave propagation. In a fig. \ref{fig2} the instant is shown when four dominoes have transferred to a state $2^{\prime}$ with a low potential energy $U^{\prime}_{2}$ ($U^{\prime}_{2}$ exceeds $U_{2}$ a little). 
Fifth domino has overcome a potential barrier and this domino matches a position of the wave front propagating from left to right. It is supposed that the domino being extreme at the left has received the energy $E_{0}$ in an initial instant ($E_{0}$ exceeds a barrier $\Delta U_{31}$). Then the further autocatalytic process develops under condition of: $E^{\prime} = \Delta U^{\prime}_{32} = U_{3} - U^{\prime}_{2} > \Delta U_{31}$. Let's note that during interaction of dominoes among themselves and with a surrounding medium the part of energy is dissipated. Therefore quantity $E^{\prime}$ should exceed $\Delta U_{31}$ not less than on quantity of dissipated energy. 

\textbf{About a wave velocity.}

It is clearly that even in an ideal case, when the energy dissipation misses, the switching wave velocity $C \longrightarrow 0$ if $\Delta U_{12} \longrightarrow 0$ as the exuberant energy for undular transport peters. Therefore the ascending of $ó$  because of  the ascending of   $\Delta U^{\prime}_{12} = U_{1} - U^{\prime}_{2}$ is natural. For the fixing of a functional connection view it is enough to use the energy conservation law. Then it is evidently that $C$ is proportional to $\sqrt{\Delta U^{\prime}_{12}}$.
Dependence of the starting and the wave propagation from the domino's sizes relation Let's consider that a mass of a domino is constant. Then the diminution of the basis size together with the ascending of a vertical dimension (in the first equilibrium position) are accompanied by ascending of $U_{1}$ by decrease of $U_{2}$ and accordingly by ascending of velocity $C$. 
Similar change of the form besides leads to the decreasing of an energy barrier $\Delta U_{31} = U_{3} - U_{1}$ and to the decreasing of the stability of an initial state with energy $U_{1}$. Then the energy $E_{0}$, which is necessary to the switching wave beginning, also is reduced. If the homogeneous system of a dominoes has very small energy of a barrier a place of the switching wave beginning is casual.
Such place is oozed by small diversions (by fluctuations) from medial values of parameters of a surrounding medium. As casual diversions to the left and to the right are equality probability, in system of elements with a low energy threshold there can be the switching waves with opposite directions of propagation. 

\textbf{About global and local infringement of symmetry.}

\begin{figure}[htb]
\centering
\includegraphics[clip=true, width=0.8\textwidth]{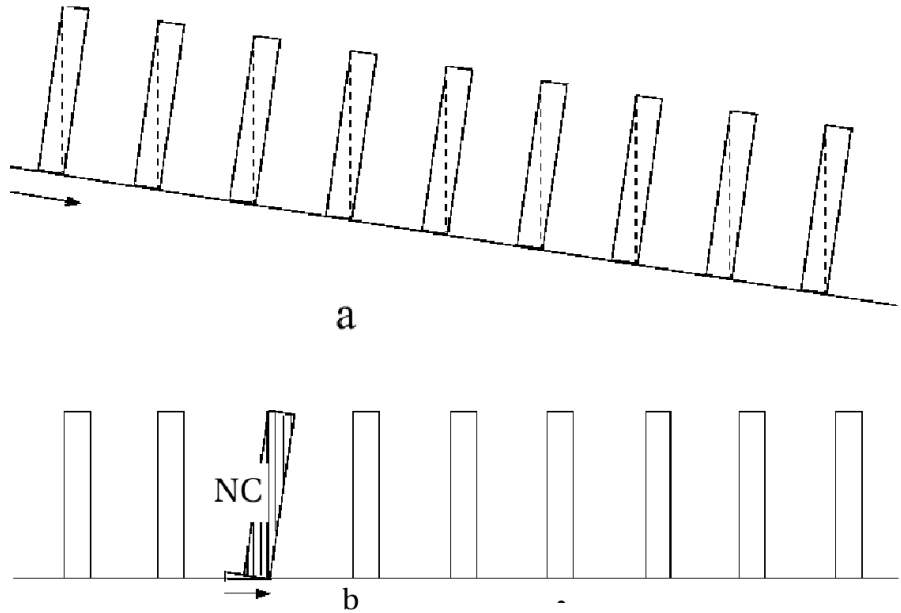}
\caption{Infringement of symmetry: a) global; b) local.}
\label{fig3}
\end{figure}

If not to change the form of a dominoes the decrease of size of a barrier can be achieved at slightly rejecting from horizontal orientation of a basic plane. Then, as it is obvious from fig. \ref{fig3}a, the wave propagation direction (downwards along an oblique plane), corresponding to the overcoming of a small energy barrier, is explicitly chosen. 
Introduction of a plane declination can be interpreted as the external field influence leading to a global infringement of the diversion symmetry (to the left and to the right) from a standing of equilibrium (for all elements). Now it is easy to represent the system with the local infringement of symmetry (see fig. \ref{fig3}b.) It is clear that the chosen (canted) domino has the greatest energy and corresponds to the inhomogeneity in system (i.e. corresponds to an interior local field), leading to the switching wave in a direction indicated by an arrow. ("NC" is an abbreviation of words "nucleation centre"). 

\textbf{The intermediate resume.} 
\begin{enumerate}
\item{ The switching wave is spreading with the help of go-ahead mechanism of an energy transmission (from an element to an element).}
\item{The energy liberation begins after overcoming an energy barrier and is localized in the wave front region.}
\item{An energy barrier can vary both globally and locally due to action of an exterior fields and an internal fields. The analysis of an internal fields allows to find the most probable place of a wave start and a direction of its propagation.}
\item{The wave propagation speed is increasing if a difference of energies is increasing for states between which the switching is taking place.}
\end{enumerate}

\section{NUCLEATION AND GROWTH OF THE MARTENSITIC CRYSTAL}

In the distributed mediums in some cases it is possible to specify the spatial sizes (scales) of those meshes which play a role of effective bistable elements.
As a rule similar cells contain a lot of microparticles and there is a problem about the mechanism providing the coordinated overcoming of an energy barrier by microparticles. 

\textbf{Vivid example of such behaviour is the growth of martensitic crystals in iron based alloys.} 

The reconstructive phase transitions, to which the $\gamma-\alpha$ martensitic transformation (MT) is related, demonstrate pronounced features of the first-order transitions, namely, considerable temperature hysteresis (between the direct and reverse transformations) and thermal and volume effects.

\textbf{ Processes of martensite nucleation are heterogeneous.}

\begin{figure}[htb]
\centering
\includegraphics[clip=true, width=0.8\textwidth]{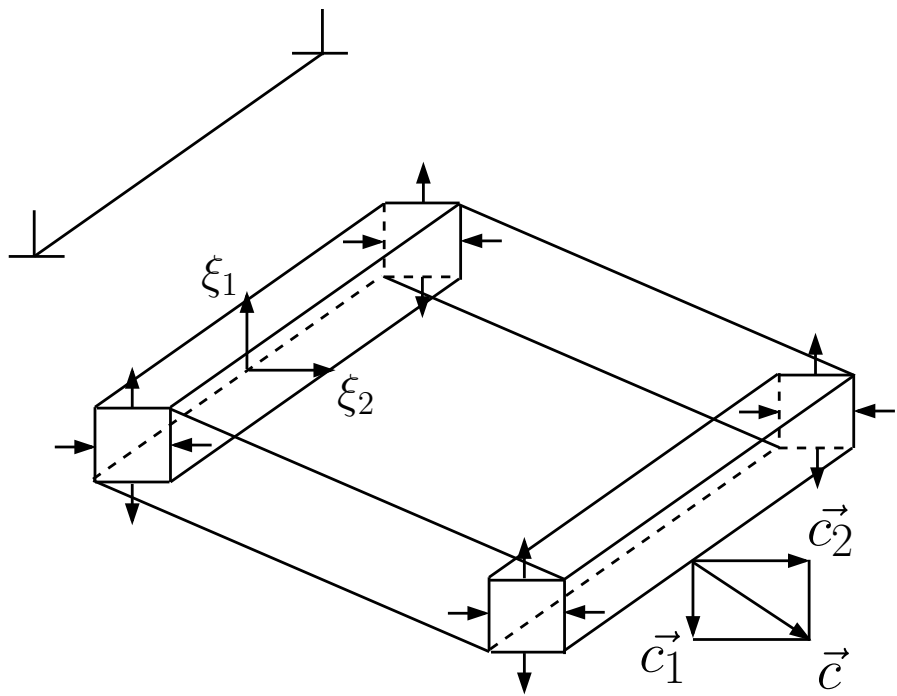}
\caption{Nucleation and growth of martensitic crystal in the model of the modified switching wave.}
\label{fig4}
\end{figure}

To the features of MT it is necessary to refer a series of morphological characteristics and, certainly, the growth of the martensitic crystals with the hyper sound speed being the major feature of a microkinetics. The last feature is showing the undular nature of the mechanism controlling the growth of the martensitic crystals.
Features of MT are explained in the switching wave model with quite concrete dynamic oscillatory structure of the transition (interphasic) region being a front of the switching wave. 
Now we will list the key standings of model (the dynamic distinctive features of model are submitted on fig. \ref{fig4}).
\begin{enumerate}
\item{The specific volume fulfills the function of the leading order parameter. The switching is taking place between values of the specific volume.} 
\item{The part of a total energy which depends from an order parameter and has two stable states is considered. The relative stability of states (as well as an energy barrier parting their) depends on temperature. In model of a domino of such dependence it is possible to correlate with change of a relation of the domino's sizes.}
\item{Now the role of a separate domino overcoming an energy barrier is played by a three-dimensional mesh in the shape of a prolate rectangular parallelepiped (see fig. \ref{fig4}).

Its pairs of edges oscillate in opposed phases. In result the flat strain of the tension-compression type will be stimulated. The excited parallelepiped radiates the wave-beames, propagating in orthogonal directions with velocities $\mathbf{C}_{1}$ and $\mathbf{C}_{2}$.}
\item{The lattice sequentially loses a stability, shaping plate-like area there, where the wave beams boosting threshold strain are imposed.}
\item{The energy for autocatalytic undular process is oozed only in the region which has tested threshold strain. Therefore hyper sound velocity $\mathbf{C}$ (the vector total of velocities $\mathbf{C}_{1}$ and $\mathbf{C}_{2}$ of wave beams) becomes the actual growth velocity of martensitic crystals. It is important to underline that the dynamic structure of wave-front is precisely the reason which leads to the hyper sound velocity of a modified switching wave. Let's note that the velocity of the unmodified wave of switching depends on a difference of specific free energies of phases \cite{bib:1} and does not reach a speed of sound.}
\item{The initial state in the shape of an oscillating parallelepiped arises in elastic fields created by defects. The defects are identified as the dislocations (on a fig. \ref{fig4} the segment restricted by symboles $\perp$ matches a dislocation line ). The parallelepiped has the edges with normals oriented along the eigenvectors ($\xi_{i}$) of the strain tensor of elastic fields created by defects. Thus directions of propagation of wave-beams are genetically connected with an elastic field of defect in the region of a nucleation.}
\item{In the stage of rapid martensite crystal growth the intensive electron currents exists in a region between the phases. This region is characterized by strong gradients of chemical potential $\mathbf{\nabla} \mu$. The gradients in turn are connected with the difference of phase specific volumes.}
\item{The part of a kinetic energy of an electron current is conversed to the energy of wave-beams owing to the generation mechanism of waves by non-equilibrium electrons. The essence is simple. There are more electrons moving against the gradient than along the gradient i.e. there are the pairs of electronic states with inverted population. Therefore the generation of elastic waves by non-equilibrium electrons is interpreted as a phonon maser effect.}
\end{enumerate}

\section{INFLUENCE OF THE FINITE DEFORMATIONS ON NUMBER OF PAIRS $R_{ef}$ OF ELECTRONIC STATES WITH INVERTED POPULATION}

\begin{figure}[htb]
\centering
\includegraphics[clip=true, width=0.8\textwidth]{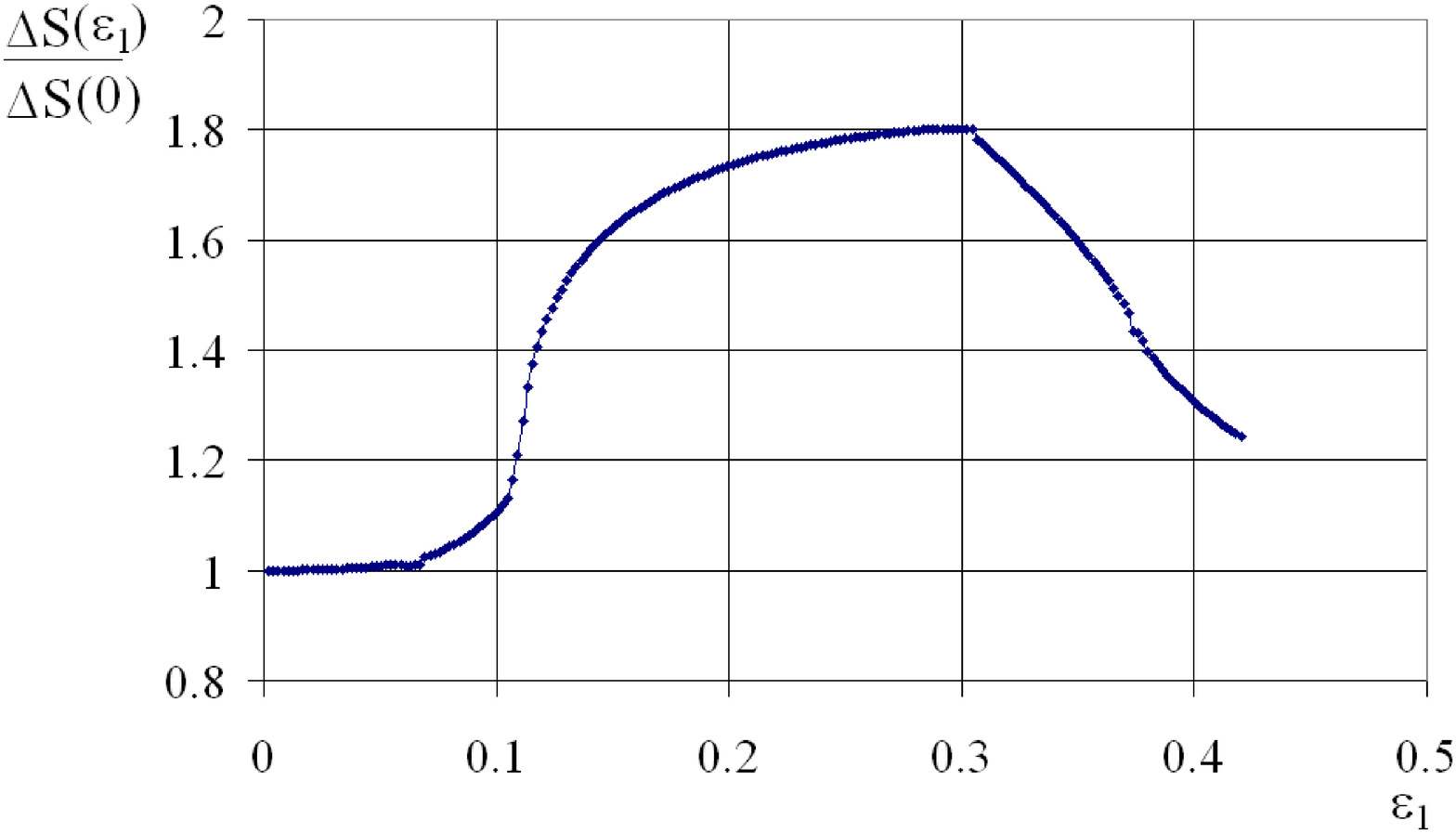}
\caption{An example of the dependence of $\Delta S(\varepsilon_{1})/\Delta S(0)$ on tensile deformation.}
\label{fig5}
\end{figure}

As shown in the previous papers generalized in the monography \cite{bib:1} in case of an ideal lattice the number $R_{ef}$ is sufficient for generation of waves carrying the threshold strain of the order of limit of elasticity $\varepsilon_{e}$. It is important to fix how the finite strain will influence onto number of $R_{ef}$. Let's remind that the lattice strain (Bein's strain) of the order of $0,1$ i.e. is equal $(10^{2}-10^{3})\varepsilon_{e}$.
The calculation of Ref was carried out for a model electronic spectrum in a tight-binding approximation \cite{bib:2}. The strain $\varepsilon$ had the same phylum as threshold strain. It is shown that there is a considerable gamut of strains, in which the value of $R_{ef}$ is increasing at the ascending of $\varepsilon$.
For example on fig.\ref{fig5} the dependence on tensile strain (along an axis of symmetry of the fourth order initial fcc lattice) of $\Delta S(\varepsilon_{1})/\Delta S(0)$ is demonstrated for an energy interval $0,6 \dim 1,0$\,eV near of Fermi-level $\mu = 0,8$\,eV (without taking into account of the dependence of $\mu$ on $\varepsilon$). 
The analysis shows that the essential increment of $\Delta S(\varepsilon_{1})/\Delta S(0)$ in the strain interval $0,06 \dim 0,4$ (the greatest growth of rate of $\Delta S(\varepsilon_{1})/\Delta S(0)$ is achieved in narrow region $0,103 < \varepsilon_{1} < 0,116$) is caused by the states located on square planes of Brillouin zone (orthogonal to $\mathbf{\nabla} \mu$). The area of this square planes increases during deformation. At a standing of Fermi-level near of the peak of a density of states the similar dependences take place and for a plane deformation. The considered examples testify about an opportunity of essential decrease of threshold value of inverted population difference $\sigma_{th}$ during deformation.

This conclusion is important for the description of a stage of growth of the crystal of martensite (when the lattice experiences the significant deformation after loss of stability) as well as  for the description of the nucleation stage of martensitic crystals. It is gives additional arguments in favor of the mechanism of rigid excitation of initial fluctuations at reconstructive martensitic transformations ($\sigma_{th}$ is decreasing when the amplitudes of fluctuations are increasing). 

It is possible to assume, that most intensive martensitic transformations take place in alloys of such composition for which the decreasing of $\sigma_{th}$ is typical during increasing deformation.

\section{CONCLUSION}

It has been shown that the interphase region is capable to generation of waves and during a plastic strain breaking symmetry of a lattice. It is important for improvement of the physical nature of a modified switching wave.

\bibliographystyle{unsrt}

Let's note that in \cite{bib:2} instead of figure 2 it is necessary to mean figure \ref{fig5} of this paper.

\end{document}